\documentclass[12pt,twocolumn]{article}
\usepackage[
backend=biber,
style=phys,
sorting=none
]{biblatex}
\usepackage{amssymb}

\newcommand{\e}{\varepsilon}

\newcommand{\G}{\Gamma}

\newcommand{\D}{\Delta}
\newcommand{\s}{\sigma}
\newcommand{\m}{\mu}

\newcommand{\up}{\uparrow}
\newcommand{\down}{\downarrow}

\newcommand{\unit}[1]{ \ \mathrm{#1}} 

\usepackage{color}
\usepackage{braket}
\usepackage{mathtools}
\usepackage{amsmath, cuted}
\usepackage{nccmath}
\usepackage{placeins}
\usepackage{hyperref}
\usepackage[hyperref]{xcolor}
\definecolor{link}{RGB}{0,128,172}
\hypersetup{
    colorlinks=true
}
\newcommand{\pt}[1]{ {\color{black}{#1}}}

\newcommand{\rev}[1]{ {\color{black}{#1}}}

\title{Thermal Generation of Spin Current in a Quantum Dot Coupled to Magnetic Insulators}
\author{Emil Siuda$^{1\dagger}$
\and
Piotr Trocha$^1$}
\date{$^1$Institute of Spintronics and Quantum Information, Faculty of Physics, Adam Mickiewicz University\\61-614, Poznań, Poland\\
$\dagger$ \url{emisiu@amu.edu.pl}
}

\begin{document}
\maketitle
\hypersetup{
  linkcolor=black,
  urlcolor=link,
  citecolor=black
}


\section{Introduction}\label{sec:Introduction}
Recent three decades reveal extensive studies on utilizing the spin of an electron alongside its charge as an information carrier in electronics within the relatively new branch of solid-state physics called spintronics \cite{guo2021spintronics}. During this effort, it turned out that magnons, being quanta of spin waves, can be excellent carriers of spin current, giving rise to the emergence of magnonics \cite{barman20212021}. The advantages of magnonic spin current over the electronic one include negligible Joule heating, low energy consumption, long spin diffusion length/mean free path, and efficient tunability with a number of parameters \cite{barker2016thermal,pirro2021advances}. Electronic devices based on magnons, such as logic gates, diodes, directional couplers, memories, and others, have already been proposed \cite{borlenghi2014designing, gertz2014magnonic, miura2018tunable, goto2019three, talmelli2020reconfigurable, wang2020magnonic, song2021spin}.

Magnon current can be driven by various means. One promising approach is based on the spin-thermoelectric effects \cite{bauer2012spin}. Exploiting a temperature gradient as a driving force allows changing the Joule heat generated, for instance, by conventional electronic devices, which is normally emitted to the environment, into useful energy. This makes thermoelectric devices environmentally friendly. 

\pt{Spin current can be driven by various means including inductive microwave technique \cite{schneider2008phase}, ultra-short laser pulses \cite{lenk2011building}, electrically 
by spin-transfer torque technique \cite{demidov2010direct, madami2011direct, demidov2012magnetic}, passing charge current through metallic films deposited on magnetic insulator \cite{cornelissen2015long, jeon2014voltage} or sending spin-polarized charge current through a ferromagnetic tunnel contact \cite{jeon2015relative}.}
\pt{ In turn, many proposals on thermal spin-current generators and converters have been implemented experimentally, including a  spin Seebeck effect (SSE) in magnetic insulator with attached metallic films \cite{uchida2010spin,meier2015longitudinal}, SSE in a non-magnetic semiconductor \cite{jaworski2012giant}, longitudal SSE in layered ferromagnetic insulators covered by metallic films \cite{ito2019spin}, thermal spin current from a ferromagnet to silicon induced by Seebeck spin tunnelling \cite{le2011thermal}. Moreover, it has been shown that thermal spin current in magnetic tunnel contacts to semiconductors can be fully controlled electrically \cite{jeon2014voltage}. Apart from that, it has been observed that the thermal generation
of spin current in a ferromagnetic tunnel contact on a semiconductor is more efficient than electrical one \cite{jeon2015relative}. Spin current can also be controlled by valve effect in insulating magnon junctions~\cite{guo2018magnon,guo2020nonlocal,he2021magnon}. Theoretical research within the linear response formulation provides a qualitative and quantitative understanding of magnon-driven SSE in  ferromagnetic insulators~\cite{adachi2011linear}. In turn, the developed 
 transport theory of diffusive spin and heat transport by magnons in magnetic insulators with metallic contacts~\cite{cornelissen2016magnon} allowed to obtain 
results for the spin Seebeck coefficient in YIG with Pt contacts which agree with the published experiment \cite{cornelissen2015long}. Generation of pure spin current has been  predicted in thermally driven molecular magnetic junctions~\cite{wu2019pure}, and rectification of thermal spin current has been theoretically demonstrated in metal-magnetic insulator interfaces \cite{ren2013predicted,tang2018rectifying} or insulating magnetic junctions with localized spin~\cite{ren2014nanoscale}.}

However, quantum dot-based thermoelectric devices seem to be very promising due to their potentially high efficiency of converting heat into useful (electric and) spin current(s) \cite{sothmann2012magnon, ren2013theory, karwacki2015magnon, prete2019thermoelectric, trocha2022spin}. Thus, in the present work, we investigate thermally induced spin current in a system consisting of a quantum dot attached to external ferromagnetic insulators.
Recent investigations have shown that there is interest in the search for quantum dot systems coupled to magnetic insulators \cite{jin2020synthesis}.

Most studies concerning the transport of magnons in such systems assumed energy-independent coupling between the quantum dot and the magnonic reservoirs \cite{sothmann2012magnon, karwacki2015magnon, trocha2022spin}, and only a few papers consider the energy-dependent magnonic density of states \cite{ren2013theory, siuda}. Generally, in magnonic systems, the wide band approximation - usually valid in fermionic systems - is not valid, and energy dependence of the density of states has to be explicitly considered. To meet this requirement, here, we consider the energy-dependent density of states of the magnonic reservoirs. This assumption leads straightforwardly to energy-dependent couplings between the dot and external magnonic reservoirs. Our results show that the energy dependence of the couplings strongly influences the magnon current flowing through the system composed of a quantum dot coupled to two magnetic insulators. Apart from that, we also consider magnon-magnon interactions and study their impact on the thermally generated spin current of the magnonic type. \rev{These interactions are ever-present in magnetic materials and may significantly affect generated magnon current, especially in high temperatures.}\pt{Proper modelling of the magnonic density of states and inclusion of many-body interactions are crucial from the experimental point of view as shown in Ref.~\cite{rezende1,rezende2}  in which theoretical predictions, including magnonic spin current induced by the temperature gradient in a ferromagnetic insulator, have been compared to experimental data and revealing good agreement.}

Although, experimental implementation of the ferromagnetic insulator – QD –  ferromagnetic insulator is challenging, it can be realized by utilizing two-dimensional electron gas properly restricted by metallic gates \cite{kouwenhoven2001few, reimann2002electronic, hanson2007spins} and depositing additional contacts made from magnetic insulators.
Moreover as the QD filters magnons of specific energy, the proposed system can be used as a monochromator for the thermally generated spin waves. Such a possibility is inaccessible in junction structures where magnons of all available frequencies are transferred.

The paper is organised as follows. In Sec.~\ref{sec:Model and results} we present the theoretical description of the considered system. In particular, we introduce the model describing the considered system and derive the formulas for thermally generated magnon (spin) current taking into account the energy-dependence of the relevant couplings and magnon-magnon interactions in the magnonic reservoirs. The numerical results and their discussion
are presented in Sec.~\ref{sec:Results}. Finally, the paper is concluded in Sec.~\ref{sec:Conclusions}.
\section{Theoretical framework}\label{sec:Model and results}
\begin{figure*}
    \centering
    \includegraphics{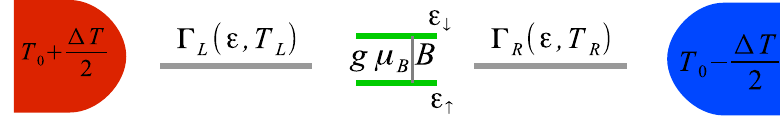}
    \caption{Schematic of the system. The quantum dot is depicted as two green lines symbolising spin-up (bottom line) and spin-down (top line) states with energy $\e_{\up(\down)}$. The difference in energies is $\D\e = g\m_B B$. The reservoirs are marked as red and blue rectangles with temperatures $T_{\alpha} = T_0 \pm \frac{\D T}{2}$. The couplings $\G_{\alpha}$ are function of the energy of magnons $\e$ and the temperature $T_{\alpha}$.}
    \label{fig:Fig1}
\end{figure*}
\subsection{Model and Hamiltonian}
The system under study is schematically depicted in Fig. \ref{fig:Fig1}. It consists of a quantum dot (QD) with a single energy level coupled to two ferromagnetic insulators treated as reservoirs of magnons. Each magnonic lead can be held at a different temperature, i.e. $T_\beta$ for the left ($\beta=L$) and right ($\beta=R$) lead. When there is a temperature difference set between the leads, the magnons flow in the direction of the difference caused by the imbalance in the occupation of the states in the left and right reservoirs. The magnon current flow is mediated via the means of the electron present on the dot. Thus, the magnon current can only flow when the dot's level is split by the Zeeman interaction with an external magnetic field $B$. This allows transitions of the electron between the spin-up and spin-down states, accompanied by emissions or absorption of the magnons with energy equal to the Zeeman splitting. Therefore, the dot filters only the magnons with energy equal to $g\mu_BB$, with $g$ denoting the dot's Lande factor and $\mu_B$ being Bohr's magneton. The Hamiltonian of the system can be divided into three parts;
\begin{equation}\label{eq:Hamiltonian}
\mathcal{H} = \mathcal{H}_\beta + \mathcal{H}_d + \mathcal{H}_t,
\end{equation}
describing the reservoirs, the dot, and the tunnelling of magnons between the two leads respectively.

The magnetic insulator is described by the Heisenberg Hamiltonian;
\begin{equation}\label{eq:Heisenberg_Hamiltonian}
\mathcal{H}_{\beta} = -J_{ex}^\beta\sum_{\left<i,j\right>\in \beta}\mathbf{S}_i\cdot\mathbf{S}_j-g_m^{\beta}\mu_BB\sum_{i\in\beta} S_i^z,
\end{equation}
where $\left<i,j\right>\in\beta$ means summation over nearest neighbors in the lattice $\beta$ ($\beta=L,R$), $J_{ex}^\beta$ and $g_m^\beta$ are the corresponding exchange integral and Lande factor. We assume the magnetic field is aligned parallel to the $z$ direction. To diagonalize Hamiltonian (\ref{eq:Heisenberg_Hamiltonian}), we perform the Holstein-Primakoff transformation \cite{primakoff1940field}, keeping terms quadratic and quartic in the magnon operators and utilizing the random-phase approximation. As a result, we obtain;
\begin{equation}\label{eq:Heisenberg_Hamiltonian_Transformed}
\mathcal{H}_\beta = \sum_{\mathbf{q}\beta}\e_{\mathbf{q}\beta}b_{\mathbf{q}\beta}^\dagger b_{\mathbf{q}\beta},
\end{equation}
with an energy dispersion $\e_{\mathbf{q}\beta} = \alpha_{\beta}\left[2Z_\beta S_\beta J^\beta_{ex}(1-\gamma_{\mathbf{q}})+g^\beta_m\mu_BB\right]$. Here, $Z_\beta$ denotes the coordination number of nearest neighbours, and $S_\beta$ stands for the spin magnitude of each atomic spin in the $\beta$th reservoir, whereas $\gamma_{\mathbf{q}} = 1/Z_\beta\sum_{\mathbf{r}} \mathrm{exp}(i\mathbf{q}\cdot\mathbf{r})$ is a geometric/structure factor dependent on the magnonic wave vector $\mathbf{q}$ and position vectors of nearest neighbours $\mathbf{r}$.

The parameter $\alpha_\beta$ takes into account the renormalization of energy dispersion due to magnon-magnon interactions and is given by \cite{bloch1962magnon};
\begin{equation}\label{mm}
\alpha_\beta=1-\frac{1}{2 J_{ex} N Z S^2} \sum_{\mathbf{q}} \frac{\rev{\bar{\e}_{\mathbf{q}\beta}}}{\exp \left[\frac{\alpha \rev{\bar{\e}_{\mathbf{q}\beta}}}{k_B T_\beta}\right]-1},
\end{equation}
where we assume that both magnonic leads have the same magnetic properties, i.e. $J_{ex}^L=J_{ex}^R$, $S_L=S_R=S$, and $Z_L=Z_R=Z$, $g_m^L=g_m^R=g_m$, \rev{$\bar{\e}_{\mathbf{q}\beta} = \e_{\mathbf{q}\beta}/\alpha$ is the energy of the magnon without renormalization} and $N$ denotes the number of primitive cells in the volume of the $\beta$th lead. It is worth noting that $\alpha_\beta$ can still be different for two reservoirs due to its temperature dependence.

For further calculations, we perform the long-wavelength expansion at $\mathbf{q}=0$ up to quadratic order, yielding;
\begin{equation}\label{eq:Dispersion_Relation}
\e_{\mathbf{q}\beta} = \alpha_\beta\frac{E_{BZ}\pi^2}{8q_{BZ}}\mathbf{q}^2,
\end{equation}
where $E_{BZ}$ and $q_{BZ}$ are the energy and the corresponding magnitude of the wave vector of magnons at the edge of the Brillouin zone of the magnonic reservoir.
\begin{figure*}[h!]
    \centering
    \includegraphics[scale=0.75]{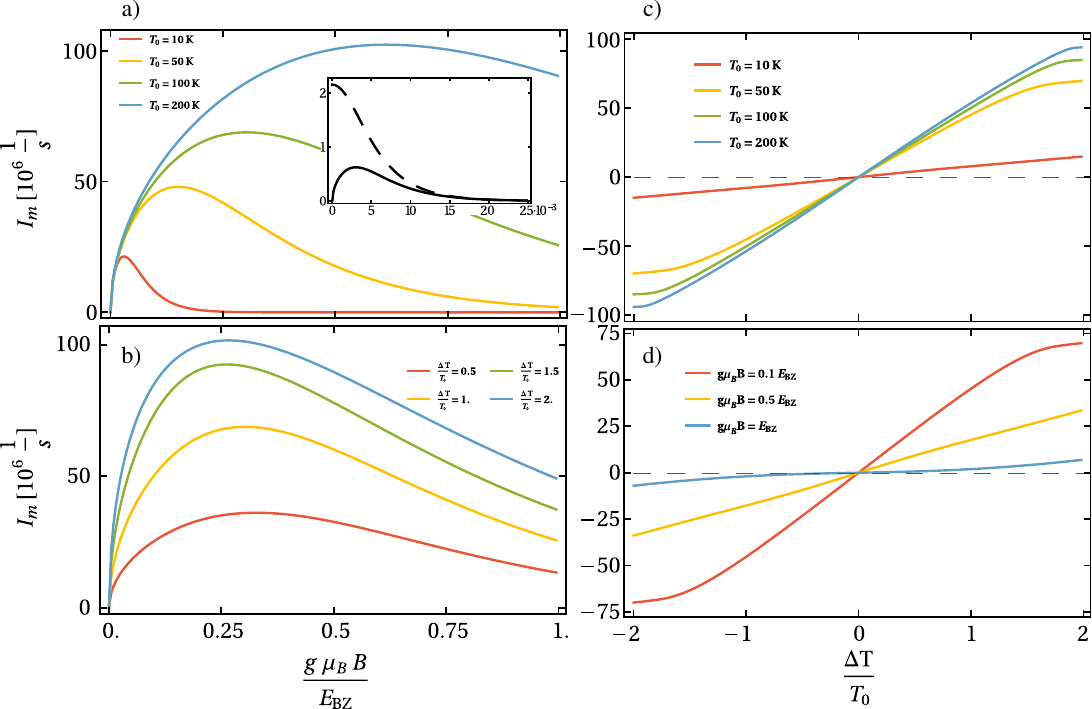}
    \caption{
        Magnon current as a function of (\textit{Left column}) applied magnetic field $B$ calculated \textbf{a)} for the indicated values of the mean temperature $T_0$ and the temperature difference $\D T = T_0$, \textbf{b)} for indicated values of the temperature difference $\D T$ and mean temperature $T_0 = 50 \unit{K}$. (\textit{Right column}) the temperature difference $\D T$ calculated \textbf{c)} for indicated values of the mean temperature and applied magnetic field $g\m_BB = 0.1E_{BZ}$, \textbf{d)} for the indicated values of the magnetic field and the mean temperature $T_0 = 50$ K. Inset presents comparison between results obtained for energy-dependent (solid line) and constant (dashed line) couplings of dot to the magnonic reservoirs calculated for $T_0 = 1$ K and $\D T = T_0$.}
    \label{fig:Fig2}
\end{figure*}

The dot is described with a single atomic level split by the Zeeman interaction induced by an external magnetic field $B$. When the dot emits (absorbs) a magnon coming to (from) the $\beta$th magnetic insulator, the electron on the dot changes/flips its spin from up to down (down to up). Due to the Pauli exclusion principle, these spin-flip processes can happen only when the dot is occupied by one electron; hence, we assume the dot is prepared in a singly occupied state and omit Coulomb interaction between two electrons occupying the dot's level, as such a situation is excluded in our model. This leads to the Hamiltonian of the form;
\begin{equation}\label{eq:Dot_Hamiltonian}
    \mathcal{H}_d = \e_{d\sigma} d_{\sigma}^\dagger d_{\sigma},
\end{equation}
where $\e_{d\s} = \e_0 \mp \frac{1}{2}g\m_BB$ with the upper (lower) sign for spin-up (down) electron on the dot. The well-defined energy levels of the dot impose the condition $\e_{\mathbf{q}} = \e_{d\down}-\e_{d\up}$, which further implies $g \geq g_m$.

The dot is coupled to the magnetic insulators with tunnelling Hamiltonian;
\begin{equation}\label{eq:Tunneling_Hamiltonian}
    \mathcal{H}_t = \sum_{\mathbf{q},\beta}j_{\mathbf{q}}^\beta b^\dagger_{\mathbf{q},\beta}d^\dagger_{\up}d_{\down} + {\rm H.c.},
\end{equation}
Where $j_\mathbf{q}^\beta$ depends on the distribution of the spins on the interface and the coupling between these spins and the electron spin on the dot. Here, we treat them as known parameters.

\subsection{Spin current}
The spin (magnon) current is calculated in the weak coupling regime within the Pauli master equation technique. We take the current flowing from the left reservoir to the dot as positive, leading to the definition of the magnon current in the left junction;
\begin{equation}\label{IML}
    I_m^L = \G_L^{IN}P_{\up}-\G_L^{OUT}P_{\down},
\end{equation}
where $\G_L^{IN} = j_L\rho_L\left(\e\right)n^+\left(\e\right) \equiv \G_L\left(\e\right)n^+\left(\e\right)$ and $\G_L^{OUT} = j_L\rho_L\left(\e\right)n_L^-\left(\e\right)\equiv\G_L\left(\e\right)n_L^-\left(\e\right)$ are the relevant transition rates and $P_\sigma$ denotes the probability of the dot being in the state occupied by an electron with spin-up  ($\sigma=\up$) or spin-down  ($\sigma=\down$).
Here and further, $n_\beta^+(\varepsilon) = \left[\exp\left(\e/k_BT_\beta\right)-1\right]^{-1}=1+n_\beta^-(\varepsilon)$ is the Bose-Einstein distribution of magnons for the $\beta$th ($\beta=L,R$) reservoir, and $\rho_\beta$ is the corresponding density of states. Moreover, $\Gamma_\beta(\varepsilon)$ denotes the coupling strength of the dot to the $\beta$th magnetic insulator. After substitutions of the above formulas for transition rates and probabilities calculated with the help of the master equation, the magnon current (\ref{IML}) becomes;
\begin{strip}
    \begin{equation}\label{eq:Magnon_current}
         I_m = \frac{\G_L\left(\e\right)\G_R\left(\e\right)\left[n_R^+\left(\e\right)-n_L^+\left(\e\right)\right]}{\G_L\left(\e\right)\left[1+2n^+_L\left(\e\right)\right]+\G_R\left(\e\right)\left[1+2n_R^+\left(\e\right)\right]}
    \end{equation}
\end{strip}

with $\varepsilon=g\mu_BB$ being the energy of transmitted magnon.
Due to the angular momentum conservation, the magnon current satisfies $I_m\equiv I_m^L = -I_m^R$. In turn, the spin current carried by magnons flowing through the left junction is given by $I_s = -\hbar I_m$ as each magnon carries $-\hbar$ momentum.

We stress out that the dot's level widths $\Gamma_{\beta}(\varepsilon)$ are functions of the energy of the transported magnons due to the energy-dependent density of states $\rho_{\beta}(\varepsilon)$. In the long-wavelength limit, when the energy dispersion is given with Eq. (\ref{eq:Dispersion_Relation}), the density of states (DOS) becomes;
\begin{align}\label{eq:Density_of_states}
    \rho_\beta\left(\e\right) &\rev{= \left(\frac{1}{2\pi}\right)^3\int d\mathbf{q}_{\beta} \ \delta\left(\e - \e_{\mathbf{q}\beta}\right)}\\ &= \frac{4\sqrt{2}q_{BZ}^3}{E_{BZ}^{3/2}\pi^5\alpha^{3/2}_\beta}\sqrt{\e}. \nonumber
\end{align}
The energy-independent factor is given as $j_{\beta}=2\pi\left<\left|j_{\mathbf{q}}^{\beta}\right|^2\right>$, where the average is taken over all the wave vectors in a given lead.

\section{Results}\label{sec:Results}
\begin{figure*}[h]
    \centering
    \includegraphics[scale=1.2]{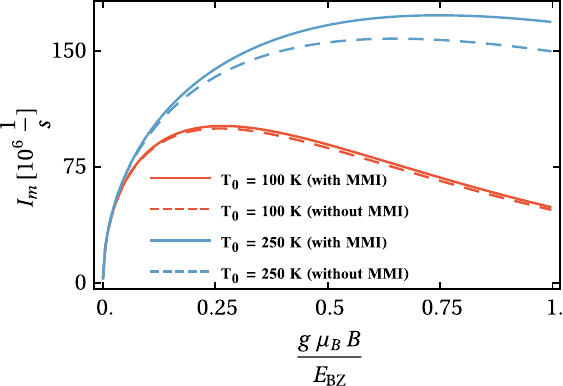}
    \caption{The influence of magnon-magnon interactions (MMI) on the magnon current. Solid lines represent the magnon current with the interactions included, while dashed lines correspond to the magnon current calculated with neglecting the interactions, setting parameter $\alpha$ to one. The temperature difference is set to $\D T = 2T_0$.}
    \label{fig:Fig3}
\end{figure*}
We assume both magnonic reservoirs are made of yttrium-iron-garnet (YIG) with the following parameters taken from the literature~\cite{rezende_book}: lattice constant $a = 1.2378$ nm, frequency of the acoustic magnons at the edge of the Brillouin zone $E_{BZ}/h = 8$ THz, magnitude of the wave vector at the edge of the Brillouin zone $q_{BZ} = 2.8/a$, exchange integral $J_{ex} = 0.82 $ meV, Lande factor $g_m = 2$, coordination number $Z = 6$, atomic spin $S = 5/2$. Moreover, we assume that $j_L = j_R = 10^{-31} \unit{eV \ m^3/s}$ which ensures weak coupling of the dot to the magnetic insulator.

Figure \ref{fig:Fig2} a) presents the magnon current as a function of the applied magnetic field for indicated values of the mean temperature $T_0$ and fixed temperature difference between reservoirs $\Delta T$. The magnon current increases sharply from zero with increasing magnetic field and achieves a maximum for some value of the magnetic field, let's say $B_{max}$. The larger the temperature $T_0$, the larger the magnetic field for which the maximum is observed.
For values greater than $B_{max}$, the current decreases monotonically to zero. This behaviour follows from the interplay between the properties of the Bose-Einstein distribution and the density of states as functions of the magnetic field. Particularly, when the applied field is low, there is an abundance of magnons, as the low temperature is enough to excite them. However, at low magnetic fields, the density of states is also low, which leads to small values of the spin current. At low temperatures, increasing the value of the magnetic field results in a fast growth of the spin current due to the rise in the density of states. After achieving the maximum, the spin current starts to drop because a lesser magnon population is available for transport even though the density of states is growing with the magnetic field. As the temperature becomes higher, the magnon current also achieves higher values. However, the maximum of the magnon current shifts to larger values of the magnetic field. This can be explained as follows: For higher temperatures, more energetic magnons can be excited, and together with the behaviour of the magnonic density of states, this explains the growth of the magnonic current and the shift of its maximum. The inset of Fig. \ref{fig:Fig2} a) compares the magnon current obtained with energy-dependent DOS [see Eq. (\ref{eq:Dispersion_Relation})] with the DOS assumed to be constant. When the DOS is taken as a constant, the low magnetic field limit $B \rightarrow 0$ of the magnon current becomes \cite{karwacki2015magnon}
\begin{equation}\label{eq:Constant_DOS_low_B_limit}
    I_m = \frac{\frac{\Delta T}{T_{0}} \G_L \G_R}{\frac{\Delta T}{T_0}\left(\G_L-\G_R\right)+2\left(\G_L+\G_R\right)}.
\end{equation}
Since the energy dependence of the DOS is neglected here, the above formula gives the maximal value of the magnon current for  $B\rightarrow 0$. \rev{This result follows from the fact, that magnons are bosonic particles and the lowest energy state is the most occupied. Since the density of states here is constant, there is no compensation to this effect similar to the one described above.} Moreover, the results for constant DOS overestimate the spin current and the magnon current is always maximal for $B$ close to zero in contrast to the current model with energy-dependent DOS.
Thus, comparing results obtained for constant density of states with the current, more general, model one notices that the former breaks in low-magnon energy regime both qualitatively and quantitatively.

In Fig. \ref{fig:Fig2} b), we show the magnon current as a function of the applied magnetic field for a given value of the mean temperature and indicated values of the temperature difference $\Delta T$. In general, increasing the temperature difference leads to an increase in the magnon current due to a greater imbalance in the occupation of the left and right magnonic reservoirs.

The magnitude of the magnon current flowing through the system is greater with a larger temperature difference between the reservoirs, as plotted in Fig. \ref{fig:Fig2} c) and Fig. \ref{fig:Fig2} d). In Fig. \ref{fig:Fig2} c), we plot the magnon current as a function of the temperature difference for the indicated values of the mean temperature $T_0$. Even though the factor $\alpha_\beta$ differs for the left and right reservoirs, the formula (\ref{eq:Magnon_current}) is symmetrical with respect to the temperature difference inversion, and thus the magnon current is too. The current is approximately linear for a quite large range of $\Delta T$. Particularly, the larger the mean temperature, the rate of increase becomes smaller at higher values of the temperature difference. However, the magnon current dependence seems to be linear for the whole range of $\Delta T$ for low mean temperatures (see red curve). Moreover, the magnon current grows with increasing values of the mean temperature.
Fig. \ref{fig:Fig2} d) presents the magnon current as a function of the temperature difference $\Delta T$ calculated for the indicated values of the applied magnetic field. It shows that the magnon current decreases with increasing applied magnetic field for the reasons mentioned in the previous paragraph.
%

Figure \ref{fig:Fig3} presents the influence of the magnon-magnon interactions on the magnon current flowing through the system. \rev{To extract the information about the interactions, we compare results obtained with $\alpha$ calculated self-consistently to ones obtained with setting $\alpha = 1$ in both leads, which indicates $\bar{\e}_{\mathbf{q}\beta}=\e_{\mathbf{q}\beta}$}. The interactions do not matter for low temperatures but become important for sufficiently high temperatures. This results from the temperature dependence of magnon-magnon interaction factor $\alpha_\beta$ given by Eq.~(\ref{mm}), which influences both the Bose-Einstein functions and coupling strengths via the energy-dependent density of states given by Eq.~(\ref{eq:Density_of_states}). Additionally, for sufficiently low magnetic fields, these interactions seem to have little impact on the magnon current, whereas, for larger $B$, they influence it greatly.
\section{Conclusions}\label{sec:Conclusions}
In conclusion, we have presented results on the magnon current in the system consisting of two magnetic insulators and a quantum dot, generated through temperature differences. Particularly, we have presented thermally generated spin current as a function of the applied magnetic field and the temperature difference between the magnonic reservoirs. We compared results obtained for constant couplings with the ones taking into account energy-dependent DOS and temperature-dependent magnonic energy renormalization. We showed that the dependency of couplings on energy is important and needs to be included, especially for low-energy magnons, to obtain correct results. Moreover, many-body interactions of magnons leave their mark on the current only for relatively high temperatures.
The presented results bring understanding of the thermally generated spin current transport in a quantum dot system coupled to magnetic insulators which may stimulate further experiments and 
help in designing spin wave monochromators.
Apart from that, the present model, including energy-dependence of the magnonic density of states, delivers more realistic predictions, than those displayed in Ref.~\cite{karwacki2015magnon}, especially in a low-energy regime, and thus, are more relevant for comparison with future experimental data.
\section*{Acknowledgements}\label{sec:Acknowledgements}
This work was supported by the National Science Centre in Poland through Project No. 2018/31/D/ST3/03965.
\FloatBarrier


\printbibliography
\end{document}